# Multilevel Resistance Switching and Enhanced Spin Transition Temperature in Single Molecule Spin Crossover Nanogap Devices

Alex Gee,[1] Ayoub H. Jaafar,[2] Barbora Brachňaková,[3] Jamie Massey,[4] Christopher H. Marrows,[4] Ivan Šalitroš,[3,5,6] N.T. Kemp[1,*]

[1]*Department of Physics and Mathematics, University of Hull, Hull, HU6 7RX, United Kingdom*

[2]*Physics Department, College of Science, University of Baghdad, Baghdad, 10071, Iraq*

[3]*Department of Inorganic Chemistry, Faculty of Chemical and Food Technology, Slovak University of Technology, 812 37 Bratislava, Slovakia*

[4]*School of Physics and Astronomy, University of Leeds, Leeds, LS2 9JT, UK*

[5]*Central European Institute of Technology, Brno University of Technology, Purkyňova 123, 61200 Brno, Czech Republic*

[6]*Department of Inorganic Chemistry. Faculty of Science, Palacký University, 17. Listopadu 12, 771 46 Olomouc, Czech Republic*

* Author to whom correspondence should be addressed. Electronic mail: N.Kemp@hull.ac.uk

**Abstract**

Spin crossover (SCO) molecules are promising bi-stable magnetic switches with applications in molecular spintronics. However, little is known about the switching effects of a single SCO molecule when it is confined between two metal electrodes. Here we examine the switching properties of a [Fe(III)(EtOSalPet )(NCS)] SCO molecule that is specifically tailored for surface deposition and binding to only one gold electrode in a nanogap device. Temperature dependent conductivity measurements on SCO molecule containing electromigrated gold break junctions show voltage independent telegraphic-like switching between two resistance states at temperature below 200 K. The transition temperature is very different from the transition temperature of 83 K that occurs in a bulk film of the same material. This indicates that the bulk, co-operative SCO phenomenon is no longer preserved for a single molecule and that the surface interaction drastically increases the temperature of the SCO phenomenon. Another key finding of this work is that some devices show switching between multiple resistance levels. We propose that in this case, two SCO molecules are present within the nanogap with both participating in the electronic transport and switching.

**Introduction**

The ability of spin crossover (SCO) compounds to exist in a bi-stable spin configuration makes them an attractive candidate for constructing new spintronic devices. SCO compounds have already been employed in the design of electromechanical actuators,[1] thermochromic displays[2] and data storage.[3] However, there remains a long way to go from current research to the ultimate goal of using these compounds as the building blocks of an electrically addressable memory technology, especially in the case of miniaturising towards the molecular scale and even single molecule devices.[4]

The SCO phenomenon has been know about since the 1930s,[5] where it was discovered that certain compounds can undergo a transition between a high spin (HS) and low spin (LS) state via the application of an external stimulus e.g. the selective wavelength irradiation, pressure or most commonly a change in temperature. In these compounds, a central metal ion is bound to ligands with an octahedral symmetry. The ligand field lifts the degeneracy of the metals ions d orbitals resulting in two sets of energy levels separated by a splitting energy (Δ), as shown in figure 1b.

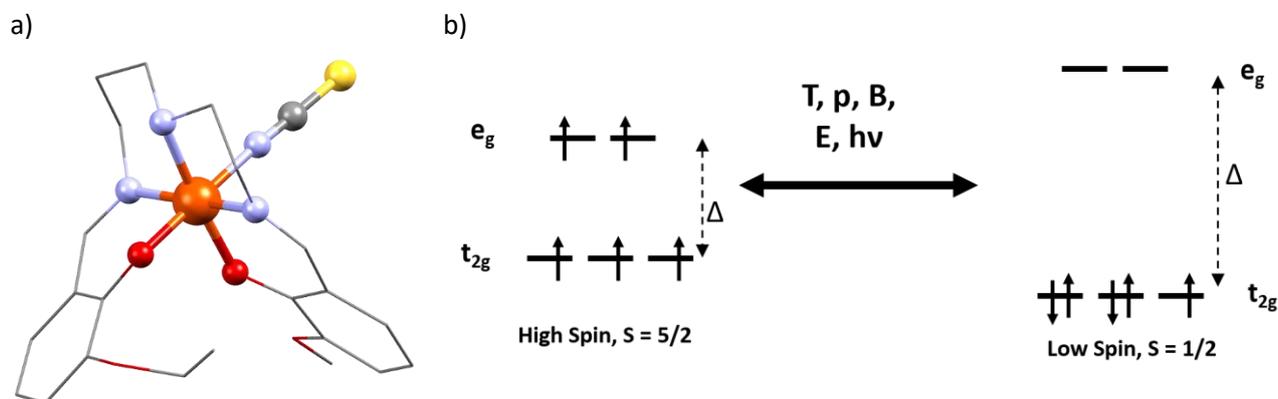

Figure 1 a) Molecular structure of the [Fe(L)NCS] complex. Fe-orange, N-blue, O-red, C-grey, S-yellow. b) Resulting energy levels for the HS and LS states split by energy Δ.

The switching mechanism is caused by a change in the interaction strength between the metal ion and ligands as a result of the external stimulus. At a certain point it becomes energetically favourable for the d electrons of the ion to either populate or depopulate the higher energy $e_g$ orbitals based on the interaction of this external stimulus. For example, an increase in temperature results in a reduced ligand-metal ion field strength due to thermal expansion increasing the bond distances which at a certain temperature results in the spin transition from LS to HS.

In recent years, there has been progress in electrically contacting these compounds at a single molecule level using scanning tunnelling microscopes (STM) and nanogap break junctions. Whilst STM has been successful in studying the underlying physics,[3,6–9] it is not suitable for real-world devices. Nanogap break junctions come in two main types. Flexible substrates that with bending, open and close the nanogap for molecular insertion. These have the advantage of controllable nanogap size,[10,11] but the approach is incompatible with current scaling techniques and with traditional semiconductor fabrication flow processes. In this study we use the electromigration method,[12–14] which has the advantage of scalability and a rigid silicon wafer substrate for better device stability. This is particularly important in the case of magnetic electrodes,[15] where SCO compounds have the potential to be used as highly efficient spin-filters,[16–18] as it can eliminate effects from magnetostriction.[19]

DFT modelling of isolated SCO compounds in contact with metallic electrodes predicts that in general there will be a change in conductance associated with the spin transition. In one case, DFT calculations on SCO compounds between metallic Au electrodes have predicted large conductance changes as much as 3000% between the HS and LS states.[20] However, the greatest challenge concerning the usefulness of these compounds is that the switching properties shown in the bulk material are not always preserved when they are deposited onto a surface in isolation. This is due to the strong surface interaction on the molecules, which can prevent the necessary structural changes needed for the spin transition to occur. Additionally, the cooperative effects of the bulk crystal[21] are no longer present in the case of an isolated molecule and so switching may be less energetically favourable. STM studies have confirmed theoretical predictions that the two spin states can have very different resistances owing to the large electronic and structural rearrangement that occurs during the spin transition. Additionally, they have confirmed the role of the interface on the switching mechanism, demonstrating how challenging the work is at the molecular level with attempts at decoupling surface interactions by placing either a thin insulating layer[3,22] or a secondary layer of SCO molecules[7] to maintain the SCO behaviour. Others have measured the low temperature signature of unpaired spins in gated molecular junctions containing SCO compounds by the presence of a Kondo peak in the I-V characteristics, and were able to control the spin transition by charging the SCO molecule via electrical gating.[23] A zero bias (Kondo) resonance is an indication of unpaired spins and a hallmark of SCO phenomenon in isolated molecules. However, this is a feature that is present only at low temperature.

In this work we investigate a SCO complex with general formula [Fe(L)NCS] compound, where L = N,N'-bis(3-ethoxy-2-hydroxybenzylidene)-1,6-diamino-3-azahexane, as shown in figure 1a. This compound belongs to a group of Schiff base complexes in which the sixth Fe(III) coordination site is taken up by a pseudohalido ligand. In the compound used in this study, this site is occupied by an NCS (isothiocyanate) group allowing the molecule to bind to Au surfaces. In the bulk this compound shows a sharp spin transition around 83 K with a small hysteresis loop of 2K (Supplementary Information). The synthesis of [Fe(L)NCS] has been previously reported.[24]

Results here contrasts with recent work that has observed conductance switching in electro-burnt graphene nanogaps containing [Fe(L')$_2$](BF$_4$)$_2$·CH$_3$CN·H$_2$O SCO compound in which they observe a switching process in the I-V characteristics that is random and independent of temperature. The authors aim of reducing molecule – electrode coupling to preserve the SCO behaviour using graphene has the disadvantage that telegraph-like switching has also been observed in clean electro-burnt junctions, which as discussed below, sheds doubt onto the origin of the switching mechanism.[25]

**Results and Discussion**

The results in figure 2 show the typical behaviour of an electromigrated Au nanogap junction that contains the SCO complex. Figure 2a shows the resistance vs temperature during cooling from 293 K to 70 K. The SCO phenomenon is evident at temperatures below 200 K, where the resistance of the device shows a clear bi-stable effect with fluctuations in the resistance between two well-defined levels. We ascribe these two resistance levels to the two different spin states of the molecule, which appears to freely switch without being triggered by light or current. We note here that during all measurements the sample was kept in the dark and the current used for measuring, 22 nA, is too small to trigger the switching. Furthermore, the induced switching of SCO complexes has been shown to be an electric field driven process, requiring upwards of 0.5 - 0.75 V in STM and mechanical break-junction devices.[3,26] At 83 K the switching effect is well established, and the lifetime of the states are long-lived with each state existing for a maximum of 16 minutes, as shown in figure 2b. Resistance histograms of the state of the molecule at 83 K and 293 K are shown in figure 2c and 2d, respectively, and clearly shown the two well-defined states at the lower temperature. Analysis of figure 2c shows that on average the molecule is in the low resistance state slightly more of the time (≈54%) than the high resistance state.

In contrast, figure 3a shows the situation when multi-level switching behaviour of a device was observed in a device after heating to room temperature a device that previously showed bi-stability. This is likely as a result of a second SCO complex either becoming active or moving into the nanogap after an extended duration at room temperature. This behaviour can be explained if instead of one molecule in the junction, there are two molecules situated in the nanogap region. In this case, the total tunnelling current is influenced by the relative contributions from both molecules. Monitoring the resistance over a long period of time a resistance histogram plot was obtained, as shown in figure 3b. Four narrow peaks are clearly visible. These are expected to be due to the following spin combinations of the two molecules within the gap (LS-LS, HS-LS, LS-HS, HS-HS). Figure 3c shows the resistance histogram of the device before cooling. In this case only a single broad peak is observed, indicating that the molecule is in the HS state due to the temperature dependence of the splitting energy, Δ. We note here that the greater broadness of this peak is due to the larger thermal noise at high temperature.

In both cases, resistance fluctuations from the structural motion of the electrode can be excluded, as this process is unlikely to present itself as a clear two or four level stability in the conduction. If this were the case, a random evolution in the measured resistance over many values or irreversible jumps in the resistance because of different molecular arrangements on the surface would be instead observed. Moreover, this behaviour would be strongly temperature dependent, being reduced at cryogenic temperatures as atomic motion is reduced. Voltage induced changes in the electrode structures can also be excluded since we limit the measurement to very small applied bias of 8 mV.[27] In devices broken without the deposition of molecules we see only a weak temperature dependence in the tunnelling resistance after cooling of the device to 70 K, as shown in Figure S3c (Supplementary Information).

There is only one previous report of similar telegraphic-like switching behaviour reported, albeit at a much lower temperature, in molecular junctions containing SCO compounds. Electro-burnt graphene devices containing a $Fe(L')_2](BF_4)_2·CH_3CN·H_2O$ compound exhibited telegraphic like conductance switching at 4 K. With the aid of DFT calculations, Burzurí *et al.* were able to show[6] that due to changes in the electronic structure of the compound, the two particular spin states of the compounds manifest as different levels of current suppression in the I-V curves with the LS state corresponding to a large current suppression and HS presenting less.

The structure of the SCO compound used here differs significantly from this previous study based on $[Fe(L')_2](BF_4)_2·CH_3CN·H_2O$. The complex used in this study, being smaller and lacking the large pyrene end groups and extended linking units do not impart such a high degree of spin-state switching sensitivity on the ligand positions that $[Fe(L')_2](BF_4)_2·CH_3CN·H_2O$ is known to have.[16] We believe that this structural difference is the key to preserving the temperature induced SCO.

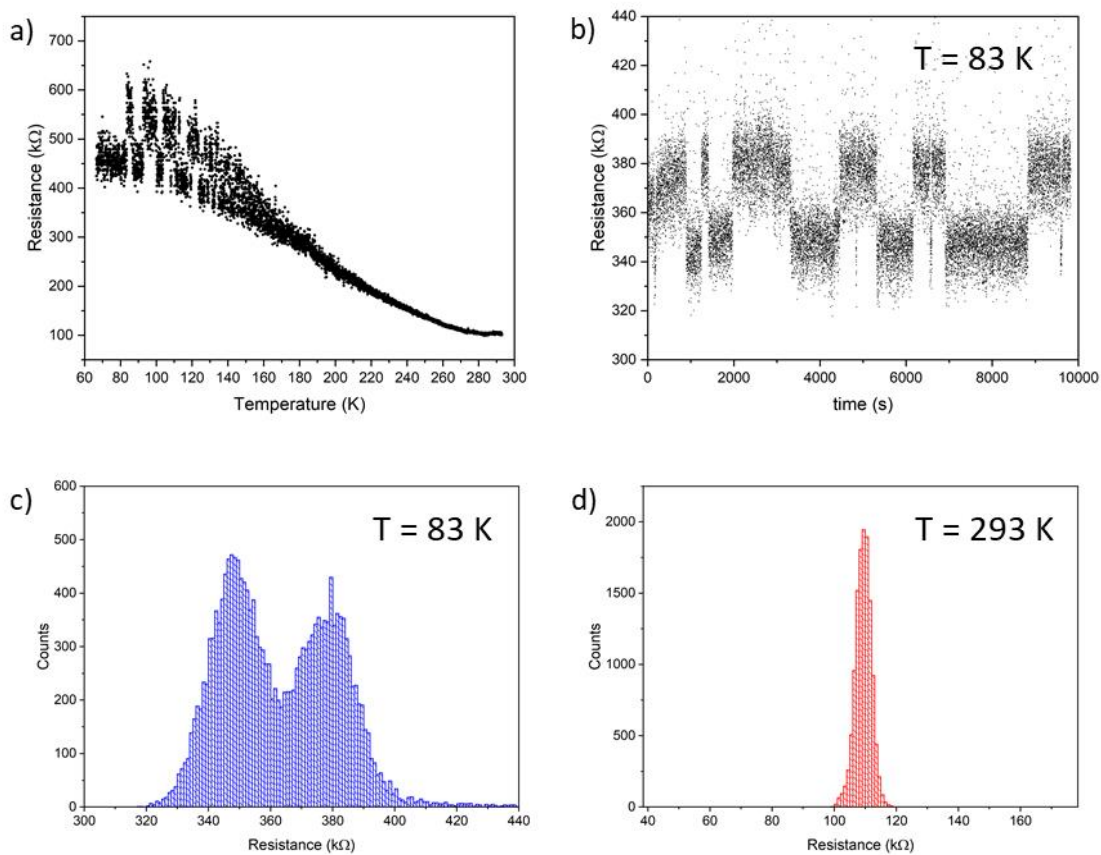

Figure 2. a) Resistance vs temperature during cooling from 293 K to 70 K for an electromigrated Au nanogap containing the SCO compound and showing the appearance of the SCO phenomenon at temperatures below 200 K. b) Resistance vs time of the device at 83 K. Resistance histograms at c) 83 K and d) 293 K. (Note: Bin size is 1 kΩ for each of the histograms with $1.2×10^4$ total samples).

The emergence of the resistance switching fluctuations between two states at temperatures below 200 K in figure 2 shows that the switching behaviour of the isolated molecule is drastically modified compared to that of the bulk material. This is not surprising since in a bulk material there is a cooperative switching effect that makes it difficult for a single molecule to switch by itself. Instead, the cooperative effect enforces all molecules to switch at once and they remain locked in that state thereafter (unless heated again above its SCO transition temperature). Furthermore, molecules in bulk films are not strongly influenced by metal surfaces since most of the molecules are not in the direct vicinity of the surface. In contrast, for a single molecule device, the molecule is attached to a metal electrode and experiences a strong coupling with the metal surface and this has been shown to drastically change the molecule's switching properties. In some cases, SCO compounds remain locked in one spin state when deposited on surfaces while other examples show the influence of a nearby surface, which can even induce HS behaviour in compounds which are otherwise fixed in a LS state.[3, 28]

In contrast to a previous report, which reported telegraphic-like switching within I-V sweeps,[16] this work used only a small bias voltage to avoid the effects of voltage triggered conformational changes[26] and resistance switching memory effects from filamentary formation.[29] Additionally, we note that telegraphic-like switching has been observed in empty graphene nanogaps resulting from interfacial defects.[25] In this study we have used clean gold nanojunctions, yet still observe telegraphic-like switching at very low and constant bias (8 mV), which is specifically used to avoid these above effects. In this work, it is more likely that the switching process arises from a reduction in the energy of the ligand field splitting parameter, Δ, as a result of the reduced temperature in conjunction with thermal perturbations of the complex. In the previous report the large extended $Fe(L')_2](BF_4)_2 \cdot CH_3CN \cdot H_2O$ molecule and its extended arrangement across electrodes is potentially more susceptible to thermal perturbations than the far smaller SCO complex used in this work.

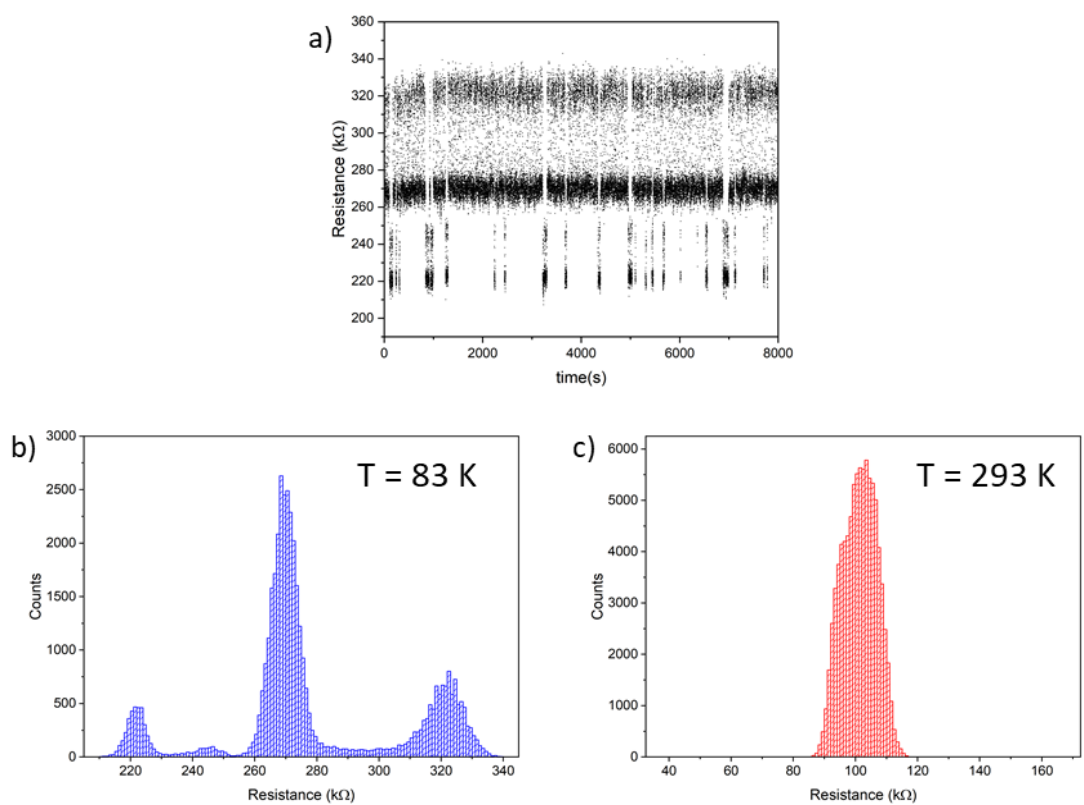

Figure 3. a) Resistance vs time for a device displaying multilevel resistance switching at 83 K. b) Resistance histograms at 83 K showing four individual peaks, which are expected to be due to the following spin combinations that occurs when two molecules are present within the gap (LS-LS, HS-LS, LS-HS, HS-HS). c)

Resistance histogram at 293 K and showing only a single broad peak. (Note: Bin size is 1 kΩ for each of the histograms with 4.3×10⁴ total samples).

**Experimental**

Devices are prepared using a two-step bilayer nanoimprint and photolithography process.[4] First, a bilayer resist structure is coated onto oxidised silicon substrates. The top layer is a low molecular weight PMMA (15k $M_w$) which is coated onto a PMGI layer. Both layers are baked for 10 minutes at 190°C to drive off residual solvent. Nanoimprint lithography is used to pattern the top PMMA layer with a stamp prepared using EBL and RIE. After a brief plasma etch the bottom PMGI layer is partially removed so that an undercut of the PMMA layer is created after development. 15nm of Au is deposited using e-beam evaporation in a system with a base pressure of 1×10⁻⁷ mbar and lift-off is performed first in acetone and then N-Methyl-2-pyrrolidone. These features are then connected to bonding pads using photolithography. A complete device containing 15 nanojunctions is shown in figure 4a.

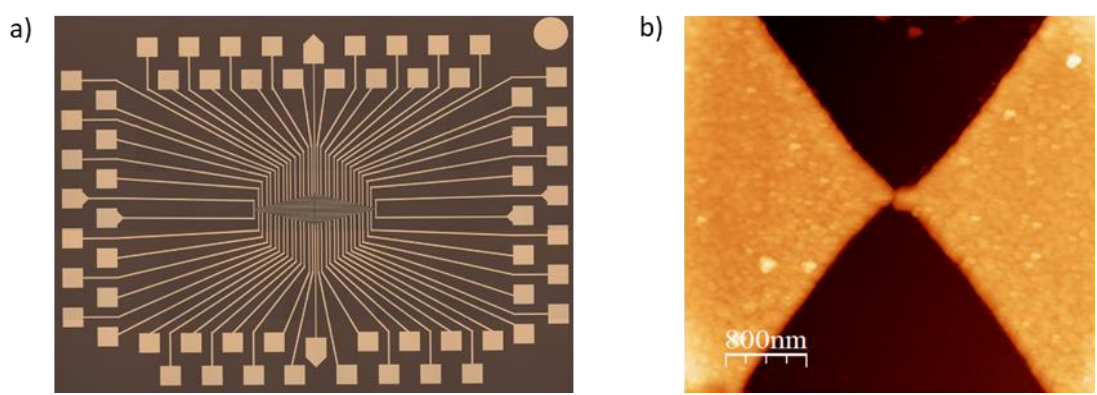

Figure 4 a) Device containing 15 nanojunctions with bonding pads. b) Nanogap after feedback controlled electromigration.

Molecules are deposited onto the nanojunctions by first oxygen plasma cleaning the devices at low power before depositing a solution of 0.01 mM [Fe(L)NCS] (see[24] for details of the [Fe(L)NCS] molecules) dissolved in acetone onto the device. After careful drying with nitrogen the devices are then mounted in a custom-made sample holder within a continuous flow cryostat and the sample space is evacuated to 1×10⁻⁷ mbar before breaking the nanojunction. A feedback controlled electromigration technique (see Supplementary Information) is used to produce a nanoscale gap in the gold nanojunction.[13,14,30] In brief, this process monitors the resistance of the nanojunction using a lock-in technique, a series of pulses of variable length and voltage are applied to the device. Through a combination of joule heating and electron-wind force the junction becomes thinned until a tunnelling gap is formed. This is indicated by a sharp jump in conductance to values well below $G_0 = e^2/h$. A feedback mechanism is used to reduce the power applied to the junction so that a thermal runaway is avoided with the results that nanogaps can be produced with high yield, figure 4b.

Temperature dependent resistance measurements are performed by allowing the cryostat to warm slowly over the course of 24 hrs. Temperature is measured using a calibrated Pt-100 situated on the sample holder and resistance is monitored every 300 ms using a lock-in technique with 8 mV excitation at 989 Hz.

**Conclusion**

Using electromigrated gold nanogap break-junction devices, we have electrically detected spin state fluctuations between spin states of a SCO compound at the single molecule level. In contrast to bulk devices containing a thin-film of the SCO molecules, switching behaviour occurs at a far higher temperature and without a cooperative switching effect. We measure a strong temperature dependence of the conductivity as well as a temperature induced spin transition that manifests in two well-defined resistance levels at low temperature, while at room temperature, only one resistance state is measured. The two well-defined resistance levels are expected to be due to conformational change in the SCO molecule as a result of thermal perturbations and the reduced energy of the ligand field splitting parameter that occurs at low temperature.

In some cases, switching between four resistance states occurred. In this case, we propose that two SCO molecules are present within the nanogap and are participating in the electron tunnelling and spin-transition process. Behaviour such as this, when controllable, could facilitate multi-level resistance switching and the development of ultra-small memory having more than two binary states.

## Conflicts of interest

There are no conflicts to declare.

## Acknowledgement

Slovak grant agencies (APVV-18-0197, APVV-18-0016, VEGA 1/0125/18) and Ministry of Education, Youth and Sports of the Czech Republic under the project CEITEC 2020 (LQ1601) are acknowledged for the financial support. I.Š. acknowledges the financial support from institutional sources of the Department of Inorganic Chemistry, Palacký University Olomouc, Czech Republic. J.M. and C.H.M. were supported by the EPSRC grant EP/M000923/1.

## References

1   I. A. Gural'Skiy, C. M. Quintero, J. S. Costa, P. Demont, G. Molnár, L. Salmon, H. J. Shepherd and A. Bousseksou, *J. Mater. Chem. C*, 2014, **2**, 2949–2955.

2   O. Kahn and C. J. Martinez, *Science*, 1998, **279**, 44–48.

3   T. Miyamachi, M. Gruber, V. Davesne, M. Bowen, S. Boukari, L. Joly, F. Scheurer, G. Rogez, T. K. Yamada, P. Ohresser, E. Beaurepaire and W. Wulfhekel, *Nat. Commun.*, 2012, **3**, 936–938.

4   A. Gee, A. H. Jaafar and N. T. Kemp, *Nanotechnology*, 2020, In press, doi.org/10.1088/1361-6528/ab6473.

5   L.Cambri, *Eur. J. Inorg. Chem.*, 1931, **64**, 2591–2598.

6   M. S. Alam, M. Stocker, K. Gieb, P. Müller, M. Haryono, K. Student and A. Grohmann, *Angew. Chemie - Int. Ed.*, 2010, **49**, 1159–1163.

7   T. G. Gopakumar, F. Matino, H. Naggert, A. Bannwarth, F. Tuczek and R. Berndt, *Angew. Chemie - Int. Ed.*, 2012, **51**, 6262–6266.

8   A. C. Aragonès, D. Aravena, J. I. Cerdá, Z. Acís-Castillo, H. Li, J. A. Real, F. Sanz, J. Hihath, E. Ruiz and I. Díez-Pérez, *Nano Lett.*, 2016, **16**, 218–226.

9   S. Ossinger, H. Naggert, L. Kipgen, T. Jasper-Toennies, A. Rai, J. Rudnik, F. Nickel, L. M. Arruda, M. Bernien, W. Kuch, R. Berndt and F. Tuczek, *J. Phys. Chem. C*, 2017, **121**, 1210–1219.

10  D. Xiang, H. Jeong, T. Lee and D. Mayer, *Adv. Mater.*, 2013, **25**, 4845–4867.

11  C. A. Martin, D. Ding, H. S. J. Van Der Zant and J. M. Van Ruitenbeek, *New J. Phys.*, 2008, **10**, 065008.

12  H. Park, A. K. L. Lim, A. P. Alivisatos, J. Park and P. L. McEuen, *Appl. Phys. Lett.*, 1999, **75**, 301–303.

13  J.-B. Beaufrand, J.-F. Dayen, N. T. Kemp, A. Sokolov and B. Doudin, *Appl. Phys. Lett.*, 2011, **98**, 142504.

14  Z. M. Wu, M. Steinacher, R. Huber, M. Calame, S. J. Van Der Molen and C. Schönenberger, *Appl. Phys. Lett.*, 2007, **91**, 11–14.

15  P. L. Popa, N. T. Kemp, H. Majjad, G. Dalmas, V. Faramarzi, C. Andreas, R. Hertel and B. Doudin, *Proc. Natl. Acad. Sci. U. S. A.*, 2014, **111**, 10433–10437.

16  E. Burzurí, A. García-Fuente, V. García-Suárez, K. Senthil Kumar, M. Ruben, J. Ferrer and H. S. J. Van Der Zant, *Nanoscale*, 2018, **10**, 7905–7911.

17  Z. Wen, L. Zhou, J. F. Cheng, S. J. Li, W. L. You and X. Wang, *J. Phys. Condens. Matter*, 2018, **30**, 105301.

18  J. Huang, R. Xie, W. Wang, Q. Li and J. Yang, *Nanoscale*, 2016, **8**, 609–616.

19  N. T. Kemp, H. Majjad, P. Lunca Popa, D. G. and B. Doudin, *ECS Trans.*, 2009, **15**, 3–10.

20  N. Baadji and S. Sanvito, *Phys. Rev. Lett.*, 2012, **108**, 217201.

21  S. Shi, G. Schmerber, J. Arabski, J.-B. Beaufrand, D. J. Kim, S. Boukari, M. Bowen, N. T. Kemp, N. Viart, G. Rogez, E. Beaurepaire, H. Aubriet, J. Petersen, C. Becker and D. Ruch, *Appl. Phys. Lett.*, 2009, **95**, 043303.

22  T. Jasper-Toennies, M. Gruber, S. Karan, H. Jacob, F. Tuczek and R. Berndt, *Nano Lett.*, 2017, **17**, 6613–6619.

23  V. Meded, A. Bagrets, K. Fink, R. Chandrasekar, M. Ruben, F. Evers, A. Bernand-Mantel, J. S. Seldenthuis, A. Beukman and H. S. J. Van Der Zant, *Phys. Rev. B - Condens. Matter Mater. Phys.*, 2011, **83**, 1–13.

24  P. Masárová, P. Zoufalý, I. N. Ján Moncol, M. G. Ján Pavlik, R. B. Zdeněk Trávníček and I. Šalitroš, *New J. Chem.*, 2015, **39**, 508–519.

25  P. Puczkarski, Q. Wu, H. Sadeghi, S. Hou, A. Karimi, Y. Sheng, J. H. Warner, C. J. Lambert, G. A. D. Briggs and J. A. Mol, *ACS Nano*, 2018, **12**, 9451–9460.

26  G. D. Harzmann, R. Frisenda, H. S. J. Van Der Zant and M. Mayor, *Angew. Chemie - Int. Ed.*, 2015, **54**, 13425–13430.

27  F. Prins, A. J. Shaikh, J. H. Van Esch, R. Eelkema and H. S. J. Van Der Zant, *Phys. Chem. Chem. Phys.*, 2011, **13**, 14297–14301.

28  A. Cini, M. Mannini, F. Totti, M. Fittipaldi, G. Spina, A. Chumakov, R. Rüffer, A. Cornia and R. Sessoli, *Nat. Commun.*, 2018, **9**, 480.

29  H. Suga, H. Suzuki, Y. Shinomura, S. Kashiwabara, K. Tsukagoshi, T. Shimizu and Y. Naitoh, *Sci. Rep.*, 2016, **6**, 1–9.

30  D. R. Strachan, D. E. Smith, D. E. Johnston, T. H. Park, M. J. Therien, D. A. Bonnell and A. T. Johnson, *Appl. Phys. Lett.*, 2005, **86**, 2005–2007.

# Supplementary Information

# Multilevel Resistance Switching and Enhanced Spin Transition Temperature in Single Molecule Spin Crossover Nanogap Devices

Alex Gee,[1] Ayoub H. Jaafar,[2] Barbora Brachňaková,[3] Jamie Massey,[4] Christopher H. Marrows,[4] Ivan Salitros,[3,5,6] N.T. Kemp[1,*]

[1]Department of Physics and Mathematics, University of Hull, Hull, HU6 7RX, United Kingdom

[2]Physics Department, College of Science, University of Baghdad, Baghdad, 10071, Iraq

[3]Department of Inorganic Chemistry, Faculty of Chemical and Food Technology, Slovak University of Technology, 812 37 Bratislava, Slovakia

[4]School of Physics and Astronomy, University of Leeds, Leeds, LS2 9JT, UK

[5]Central European Institute of Technology, Brno University of Technology, Purkyňova 123, 61200 Brno, Czech Republic

[6]Department of Inorganic Chemistry. Faculty of Science, Palacký University, 17. Listopadu 12, 771 46 Olomouc, Czech Republic

* Author to whom correspondence should be addressed. Electronic mail: N.Kemp@hull.ac.uk

**Experimental Setup**

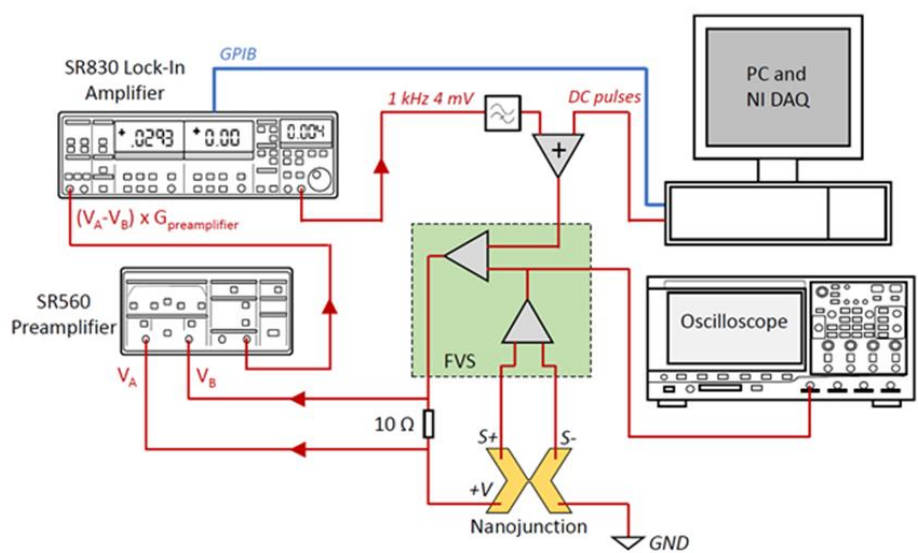

Figure S1. Four terminal, feedback-controlled setup used for electromigration of the nanojunctions.

A feedback-controlled electromigration technique is used to break gold nanojunctions forming nanogaps with separations below 2 nm. This is shown in figure S1. A pulse generator connected to a PC applies voltage pulses to the device. When the current flowing through the constriction reaches the electromigration threshold (typically $10^8$ Am$^{-2}$) the junction begins to thin. Thermal runaway is avoided in our setup using a feedback voltage source. Thermal runaway is a problem that can lead to rapid breaking and cluster formation in other electromigration schemes. The power dissipated by the nanojunction is maintained constant as the junction resistance increases preventing runaway joule heating. The junction resistance is measured via a lock-in technique. With this technique we are able to detect and monitor the progress of electromigration. With this technique we have been able to carry out electromigration smoothly to form over 70 nanojunctions.

The electrical properties of empty nanogap devices were characterized after first oxygen plasma cleaning at low power and then performing electromigration. The devices were contained in a high vacuum ($1\times10^{-7}$ mbar) throughout the experiment. Figure S2 shows the resistance of a device cooled to 70 K and allowed to warm slowly to room temperature. The resistance is sampled every 300 ms. There is no evidence of a switching process. Occasionally, a change in the resistance is observed when the device is held at room temperature for long periods of time, but this is a random process likely as a result of thermally induced diffusion of the surface electrode atoms or other structural changes in the device electrodes.

In figure S2a we show the resistance as a function of time during electromigration. The electromigration process is halted automatically when a resistance value greater than 100 kΩ is recorded. In figure S2b) the conductance is plotted in terms of $G/G_0$ (where $1/G_0 \approx 12.9$ kΩ). Conductance values take on multiples of $G_0$ before the nanojunction finally breaks, indicative of a few atom contact before the final breakage. This confirms that our electromigration process can be carried out smoothly without a thermal runaway effect that can produce metal clusters as occurs with simple electromigration processes.

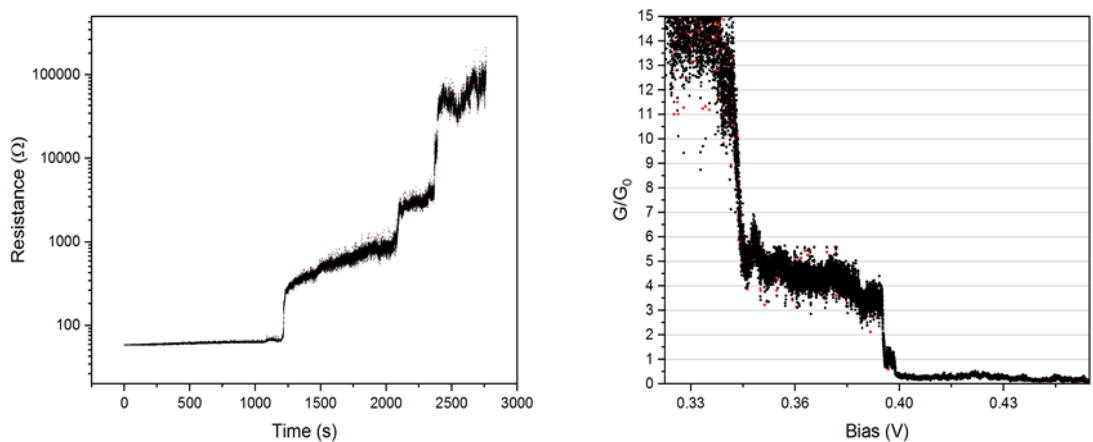

Figure S2. Characterization of a typical empty nanogap device showing a) resistance of the device as it warms slowly from 70 K to room temperature and b) conductance of the device as a function of applied bias in the region before final breakage.

## I-V characteristics of empty nanogap and temperature dependence

Figure S3 shows fits to two nanogap electrodes formed using the feedback controlled electromigration process using the Simmons equation. We find that a gap size below 2 nm fits all of the devices successfully electromigrated. Notably, the value for the work function, $\phi$, differs greatly from that of bulk gold ($\phi \approx 5.4$ eV). This discrepancy has been observed in several studies on electromigrated nanogaps[1] and is likely as a result of nanogap interfaces with structure very different from that of the regular flat gold surface with the making it highly likely that a localized non-homogeneous work function Is present in the nanogap region.[2,3,4]

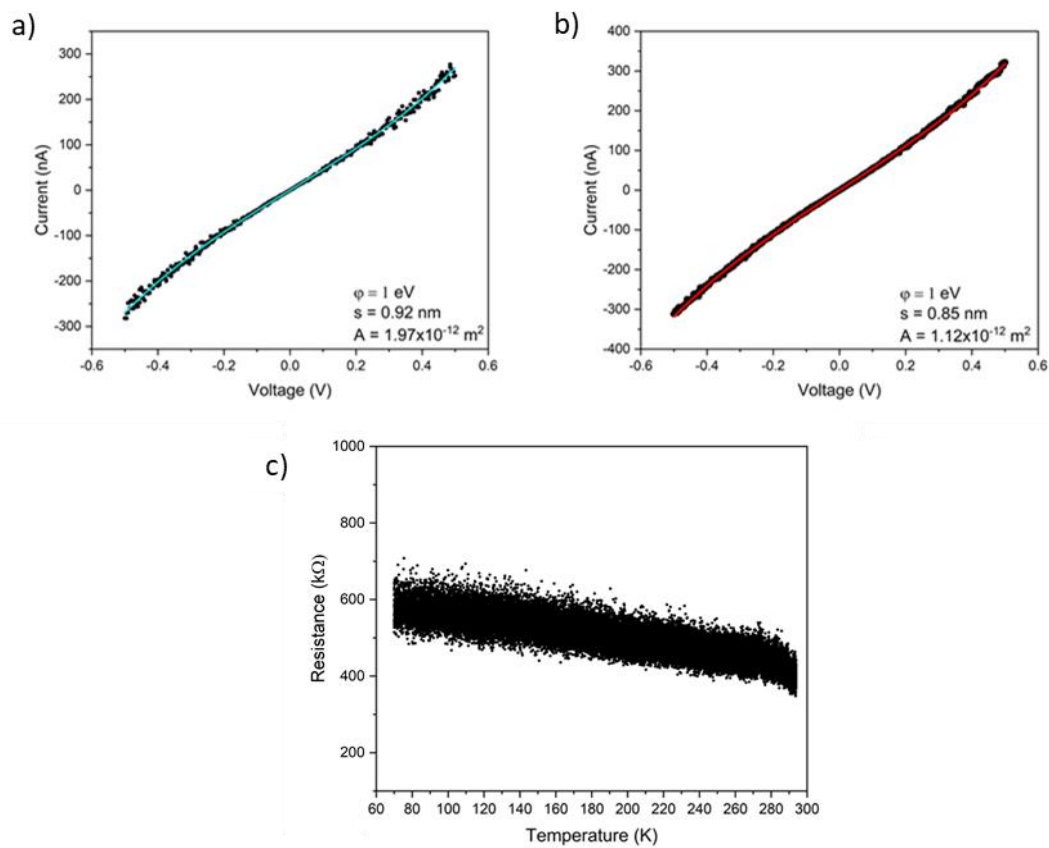

Figures S3. Fits to the Simmons model with a junction area, $A$, of $1.97 \times 10^{-12}$ m$^2$ and work function, $\phi$, of 1 eV. c) Resistance vs temperature for a clean electromigrated nanojunction measured without the deposition of molecules and broken in-situ and in high vacuum.

**Magnetic Properties of [Fe(L)NCS]**

Temperature dependent magnetometry was performed on both bulk (powder) samples as well as thin films of the [Fe(L)NCS] compound. For the thin films, the material was drop cast onto a Au thin film (50 nm thickness) on a silicon substrate. This was carried out in order to verify that the SCO behaviour is maintained when this compound is dissolved into solution and deposited as thin film directly onto a metal substrate.

A Quantum Design MPMS-SVSM system is used with an applied magnetic field of 0.1 T. Heating and cooling of the sample is carried out using the inbuilt helium flow cryostat at a rate of 3 Kmin$^{-1}$. A polypropylene capsule was used to contain the powder samples and the thin films samples were attached to a quartz paddle using GE varnish.

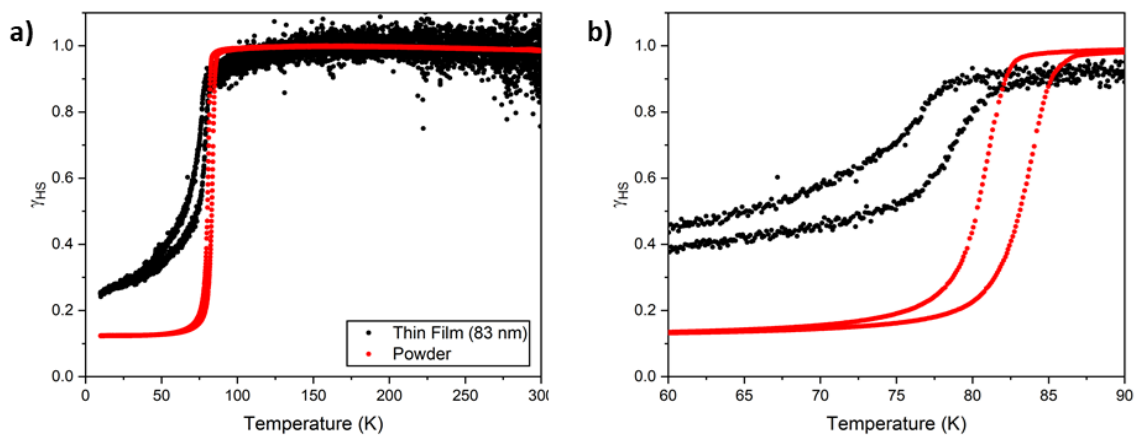

Figure S4. a) γ$_{HS}$ (high spin fraction) as a function of temperature for a powder sample (red dots) and an 83 nm thin film (black dots) of [Fe(L)NCS]. For the thin film sample, due to the low signal level and large amount of noise in the signal, the data is normalised to the mean signal value over the temperature range 300 K to 100 K. b) shows a reduced temperature range from 60 K to 90 K showing the hysteretic behaviour of the two samples.

Figure S4 shows the high spin fraction as a function of temperature for both the powder and thin film samples. A clear dependence on the sample morphology exists with both a suppression of the transition temperatures (80 K vs 68 K during cooling) and an increase in the width of the hysteresis loop (3 K compared to 7K).

Although these measurements are still probing the behaviour of a thin film and not the behaviour of a single molecule, these measurements serve to prove that our sample preparation process using drop casting from solution still preserves the spin crossover. These results qualitatively agree with magnetic studies of thermally deposited [Fe(pheny)$_2$(NCS)$_2$] films by Shi *et al.* [5] showing a rounding of the spin transition curve. It also bares resemblance to measurements made by Giménez-Marqués *et al.* for nanoparticle samples.[6] In the former study, the authors were able to prove the spin transition is preserved in samples with a thickness down to 7 nm using optical techniques. However, in order to probe the spin state behaviour of single molecules, more advanced techniques would be necessary. X-ray magnetic circular dichroism (XMCD) measurements maybe a suitable technique in order to determine the behaviour of isolated SCO complexes deposited with sub-monolayer coverage.[7,8,9]

Within the bulk SCO material, an individual switching molecule produces a localised elastic distortion within the crystal lattice. These distortions propagate through the entire material with the result being a cooperative phase transition. It is clear to see that the morphology will have some an effect on the SCO behaviour. For a thin film material, reducing the sample thickness results in a greater surface to volume ratio. Similarly, for a film comprising clusters, reducing the cluster size will again have the effect of reducing cooperativity as well as modifying the SCO transition temperature. In the extreme case, a single isolated spin crossover molecule will of course be completely absent of crystal interactions with only substrate interactions being present.

During warming and cooling processes, discontinuities were sometimes observed in the electrical resistance of the device. Figure S5 shows the resistance of a nanogap containing the SCO during a warming process from low temperature (70 K) to room temperature. In some cases, as shown, discontinuities in the resistance were observed. This is expected to be due to either changes in the structure of the electrode interface or changes within the nanogap itself, such as a movement or rotation of the attached molecule.

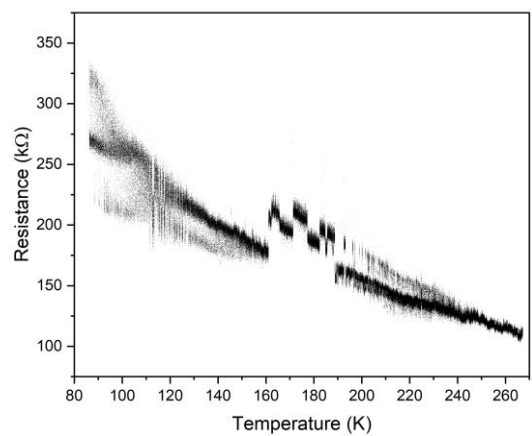

Figure S5 Nanogap containing the SCO molecules displaying multilevel switching while warming from low temperature to room temperature. The resistance was measured with a bias voltage of 8mV over a 14 hr period. In some cases, as shown, discontinuities were observed in the resistance measurement, which can be ascribed to either changes in the structure of the electrode interface or changes within the nanogap itself, such as a movement of the attached molecule.

## References

1    C. G. Wu, S. C. Chiang and C.-H. Wu, *Langmuir*, 2002, **18**, 7473–7481.

2    H. B. Heersche, G. Lientschnig, K. O'Neill, H. S. J. Van Der Zant and H. W. Zandbergen, *Appl. Phys. Lett.*, 2007, **91**, 1–4.

3    R. Hoffmann-Vogel, *Appl. Phys. Rev.*, 2017, **4**, 031302.

4    R. Schuster, J. V. Barth, J. Wintterlin, R. J. Behm and G. Ertl, *Ultramicroscopy*, 1992, **42–44**, 533–540.

5    S. Shi, G. Schmerber, J. Arabski, J.-B. Beaufrand, D. J. Kim, S. Boukari, M. Bowen, N. T. Kemp, N. Viart, G. Rogez, E. Beaurepaire, H. Aubriet, J. Petersen, C. Becker and D. Ruch, *Appl. Phys. Lett.*, 2009, **95**, 043303.

6    M. Giménez-Marqués, M. L. García-Sanz De Larrea and E. Coronado, *J. Mater. Chem. C*, 2015, **3**, 7946–7953.

7    M. Bernien, D. Wiedemann, C. F. Hermanns, A. Kru??ger, D. Rolf, W. Kroener, P. Mu??ller, A. Grohmann and W. Kuch, *J. Phys. Chem. Lett.*, 2012, **3**, 3431–3434.

8    S. Ossinger, H. Naggert, L. Kipgen, T. Jasper-Toennies, A. Rai, J. Rudnik, F. Nickel, L. M. Arruda, M. Bernien, W. Kuch, R. Berndt and F. Tuczek, *J. Phys. Chem. C*, 2017, **121**, 1210–1219.

9    B. Warner, J. C. Oberg, T. G. Gill, F. El Hallak, C. F. Hirjibehedin, M. Serri, S. Heutz, M. A. Arrio, P. Sainctavit, M. Mannini, G. Poneti, R. Sessoli and P. Rosa, *J. Phys. Chem. Lett.*, 2013, **4**, 1546–1552.